\documentclass[11pt,preprintnumbers,titlepage,nofootinbib]{revtex4-2}%
\usepackage[latin9]{inputenc}
\usepackage{multirow}
\usepackage{amsmath}
\usepackage{amssymb}
\usepackage{graphicx}
\usepackage[unicode=true,  bookmarks=false,  breaklinks=false,pdfborder={0 0
1},backref=section,colorlinks=false]{hyperref}
\usepackage{amsfonts}
\usepackage{subfigure}
\usepackage{cleveref}
\usepackage{listings}
\usepackage{xcolor}
\usepackage{array}
\usepackage{url}
\usepackage{color}%
\hypersetup{
colorlinks,linkcolor=blue,citecolor=blue,urlcolor=blue}
\makeatletter
\@ifundefined{textcolor}{}{
\definecolor{BLACK}{gray}{0}
\definecolor{WHITE}{gray}{1}
\definecolor{RED}{rgb}{1,0,0}
\definecolor{GREEN}{rgb}{0,1,0}
\definecolor{BLUE}{rgb}{0,0,1}
\definecolor{CYAN}{cmyk}{1,0,0,0}
\definecolor{MAGENTA}{cmyk}{0,1,0,0}
\definecolor{YELLOW}{cmyk}{0,0,1,0}
}

\makeatother
\begin{document}
\preprint{CTP-SCU/2022013}
\title{Observational Appearance of a Freely-falling Star in an Asymmetric Thin-shell Wormhole}
\author{Yiqian Chen$^{a}$}
\email{chenyiqian@stu.scu.edu.cn}
\author{Peng Wang$^{a}$}
\email{pengw@scu.edu.cn}
\author{Houwen Wu$^{a,b}$}
\email{hw598@damtp.cam.ac.uk}
\author{Haitang Yang$^{a}$}
\email{hyanga@scu.edu.cn}
\affiliation{$^{a}$Center for Theoretical Physics, College of Physics, Sichuan University,
Chengdu, 610064, China}
\affiliation{$^{b}$Department of Applied Mathematics and Theoretical Physics, University of
Cambridge, Wilberforce Road, Cambridge, CB3 0WA, UK}

\begin{abstract}
It has been recently reported that, at late times, the total luminosity of a
star freely falling in black holes decays exponentially with time, and one or
two series of flashes with decreasing intensity are seen by a specific
observer, depending on the number of photon spheres. In this paper, we examine
observational appearances of an infalling star in a reflection-asymmetric
wormhole, which has two photon spheres, one on each side of the wormhole. We
find that the late-time total luminosity measured by distant observers
gradually decays with time or remains roughly constant due to the absence of
the event horizon. Moreover, a specific observer would detect a couple of
light flashes in a bright background at late times. These observations would
offer a new tool to distinguish wormholes from black holes, even those with
multiple photon spheres.

\end{abstract}
\maketitle
\tableofcontents

\section{Introduction}

\label{sec:Introduction}

The Event Horizon Telescope (EHT) collaboration released images of the
supermassive black holes M87{*}
\cite{Akiyama:2019cqa,Akiyama:2019brx,Akiyama:2019sww,Akiyama:2019bqs,Akiyama:2019fyp,Akiyama:2019eap,Akiyama:2021qum,Akiyama:2021tfw}
and Sgr A{*}
\cite{EventHorizonTelescope:2022xnr,EventHorizonTelescope:2022vjs,EventHorizonTelescope:2022wok,EventHorizonTelescope:2022exc,EventHorizonTelescope:2022urf,EventHorizonTelescope:2022xqj}%
, which provides a new method to test general relativity in the strong field
regime. The main feature displayed in these images is a central brightness
depression, namely black hole shadow, surrounded by a bright ring. The edge of
black hole shadow involves a critical curve in the sky of observers, which is
closely related to some unstable bound photon orbits. For static spherically
symmetric black holes, unstable photon orbits form photon spheres outside the
event horizon. Since light rays undergo strong gravitational lensing near
photon spheres, black hole images encode valuable information of the geometry
in the vicinity of photon spheres. Therefore, black hole images have been
widely studied in the context of different theories of gravity, e.g.,
nonlinear electrodynamics
\cite{Atamurotov:2015xfa,Stuchlik:2019uvf,Ma:2020dhv,Hu:2020usx,Kruglov:2020tes,Zhong:2021mty,He:2022opa}%
, the Gauss-Bonnet theory
\cite{Ma:2019ybz,Wei:2020ght,Zeng:2020dco,Guo:2020zmf}, the Chern-Simons type
theory \cite{Ayzenberg:2018jip,Amarilla:2010zq}, $f(R)$ gravity
\cite{Dastan:2016vhb,Addazi:2021pty,Li:2021ypw}, string inspired black holes
\cite{Amarilla:2011fx,Guo:2019lur,Zhu:2019ura,Kumar:2020hgm} and other
theories
\cite{Wei:2013kza,Wang:2017hjl,Zhang:2020xub,Zeng:2021mok,Guo:2021wid,Meng:2022kjs,Qiao:2022jlu,Bogush:2022hop,Guo:2022rql,Wang:2022yvi,Zhang:2022klr,Hou:2022gge}%
.

On the other hand, testing the nature of compact objects in the universe has
been an important question in astrophysics for decades. Although the black
hole images captured by EHT are in good agreement with the predictions of Kerr
black holes, the black hole mass/distance and EHT systematic uncertainties
still leave some room within observational uncertainty bounds for black hole
mimickers. Among all black hole mimickers, ultra compact objects (UCOs), e.g.,
boson stars, gravastars and wormholes, which are horizonless and possess light
rings (or photon spheres in the spherically symmetric case), are of particular
interest since their observational signatures can be quite similar to those of
black holes
\cite{Cunha:2017qtt,Narzilloev:2020peq,Guo:2020qwk,Herdeiro:2021lwl}.
Nevertheless, it is of great importance to seek observational signals to
distinguish UCOs from black holes. For example, due to a reflective surface or
an extra photon sphere, echo signals associated with the post-merger ringdown
phase in the binary black hole waveforms can be found in various ECO models
\cite{Cardoso:2016rao,Mark:2017dnq,Bueno:2017hyj,Konoplya:2018yrp,Wang:2018cum,Wang:2018mlp,Cardoso:2019rvt,GalvezGhersi:2019lag,Liu:2020qia,Yang:2021cvh,Ou:2021efv}%
. In addition, asymmetric thin-shell wormholes with two photon spheres were
found to have double shadows and an additional photon ring in their images
\cite{Wang:2020emr,Wielgus:2020uqz,Guerrero:2021pxt,Peng:2021osd,Guerrero:2022qkh}%
. For black holes with one photon sphere, there is one shadow and one photon
ring in black hole images, and no echo signal in late-time waveforms. These
observational features would allow us to distinguish wormholes from black
holes with one photon sphere.

Intriguingly, more than one photon sphere has been reported to exist outside
the event horizon for a class of hairy black holes in certain parameter
regions
\cite{Herdeiro:2018wub,Wang:2020ohb,Gan:2021pwu,Guo:2021zed,Guo:2021ere}.
Multiple photon spheres can introduce distinctive features in black hole
images, e.g., double shadows \cite{Guo:2021ere}, extra photon rings
\cite{Gan:2021xdl} and tripling higher-order images \cite{Guo:2022muy}.
Furthermore, late-time echo signals were also observed since the effective
potential of a scalar perturbation possesses a multiple-peak structure
\cite{Guo:2021enm,Guo:2022umh}.

Can we distinguish black holes with multiple photon spheres from UCOs? To
answer this question, we investigate dynamic observations of a luminous object
freely falling in an asymmetric thin-shell wormhole in this paper. Lately,
observational appearances of a star freely falling onto black holes with a
single or double photon spheres have been numerically simulated
\cite{Cardoso:2021sip,Chen:2022qrw}. Particularly, the total observed
luminosity fades out exponentially with a declining tail, which is caused by
photons orbiting around the photon sphere, in the single-photon-sphere case.
In contrast, when there exist two photon spheres, the total luminosity
exhibits two exponential decays and a sharp peak between them. In addition,
due to photons trapped between two photon spheres, a specific observer can
detect one more cascade of flashes in the double-photon-sphere case.

Recently, luminous matter falling onto a black hole has been reported to occur
periodically near the Cyg X-1 \cite{Dolan:2011vz} and the Sgr A* source
\cite{GRAVITY:2020lpa,GRAVITY:2020xcu}. Moreover, a new way to measure the
spin of Sgr A* was proposed by simulating an infalling gas cloud
\cite{Moriyama:2019mhz}. In practice, detecting photons circling around photon
spheres several times at late times could be a challenging task due to the
scarcity of these photons. Interestingly, it showed that precise measurements
of photon rings, which are formed of photons circling around photon spheres
more than once, may be feasible with a very long baseline interferometry
\cite{Johnson:2019ljv,Himwich:2020msm,Gralla:2020srx}. Therefore, it is timely
to study observational appearances of a freely-falling star in the wormhole
background, which provides a new way to detect wormholes.

The rest of the paper is organized as follows. In Section \ref{sec:Setup}, we
briefly review the asymmetric thin-shell wormhole and introduce our
observational settings. Numerical results are presented in Section
\ref{sec:Numerical results}. Finally, we conclude with a brief discussion in
Section \ref{sec:CONCLUSIONS}. We set $G=c=1$ throughout this paper.

\section{Setup}

\label{sec:Setup}

As introduced in \cite{Visser:1989kg,Wang:2020emr,Peng:2021osd}, an asymmetric
thin-shell wormhole has two distinct spacetimes, $\mathcal{M}_{1}$ and
$\mathcal{M}_{2}$, which are glued together by a thin shell at its throat. The
metric of the wormhole is described as
\begin{equation}
ds_{i}^{2}=-f_{i}(r_{i})dt_{i}^{2}+\frac{dr_{i}^{2}}{f_{i}(r_{i})}+r_{i}%
^{2}d\Omega^{2}, \label{eq:WHmetric}%
\end{equation}
where $i=1$ and $2$ indicate quantities in $\mathcal{M}_{1}$ and
$\mathcal{M}_{2}$, respectively. Focusing on the Schwarzschild spacetime, we
have
\begin{equation}
f_{i}(r_{i})=1-\frac{2M_{i}}{r_{i}}\text{ for }r_{i}\geq R,
\end{equation}
where $M_{i}$ are the mass parameters, and $R$ is throat radius. Without loss
of generality, we set $M_{1}=1$ and $M_{2}=k$ in the rest of this paper. For
more details of the asymmetric thin-shell wormhole, refer to
\cite{Wang:2020emr}. In $\mathcal{M}_{1}$ and $\mathcal{M}_{2}$, the local
tetrads are
\begin{equation}
\mathbf{e}_{t_{i}}=f_{i}^{-\frac{1}{2}}(r_{i})\frac{\partial}{\partial t_{i}%
}\text{, }\mathbf{e}_{r_{i}}=f_{i}^{\frac{1}{2}}(r_{i})\frac{\partial
}{\partial r_{i}}\text{, }\mathbf{e}_{\theta_{i}}=\frac{1}{r_{i}}%
\frac{\partial}{\partial\theta_{i}}\text{, }\mathbf{e}_{\phi_{i}}=\frac
{1}{r_{i}\sin(\theta)}\frac{\partial}{\partial\phi_{i}}.
\label{eq:local tetrads}%
\end{equation}
At the throat, one has $\mathbf{e}_{t_{1}}=\mathbf{e}_{t_{2}}$, $\mathbf{e}%
_{r_{1}}=-\mathbf{e}_{r_{2}}$, $\mathbf{e}_{\theta_{1}}=\mathbf{e}_{\theta
_{2}}$ and $\mathbf{e}_{\phi_{1}}=\mathbf{e}_{\phi_{2}}$, which yields the
relations between the bases of the tangent space of $\mathcal{M}_{1}$ and
$\mathcal{M}_{2}$,
\begin{equation}
\frac{\partial}{\partial t_{1}}=Z^{-1}\frac{\partial}{\partial t_{2}}\text{,
}\frac{\partial}{\partial r_{1}}=-Z\frac{\partial}{\partial r_{2}}\text{,
}\frac{\partial}{\partial\theta_{1}}=\frac{\partial}{\partial\theta_{2}%
}\text{, }\frac{\partial}{\partial\phi_{1}}=\frac{\partial}{\partial\phi_{2}%
}\text{,} \label{eq:connect}%
\end{equation}
where$\;Z\equiv\sqrt{f_{2}(R)/f_{1}(R)}$. Therefore, the components of a
vector at the throat in $\mathcal{M}_{1}$ and $\mathcal{M}_{2}$ are related
by
\begin{equation}
V^{t_{1}}=ZV^{t_{2}}\text{, }V^{r_{1}}=-Z^{-1}V^{r_{2}}\text{, }V^{\theta_{1}%
}=V^{\theta_{2}}\text{, }V^{\phi_{1}}=V^{\phi_{2}}.
\label{eq:connect-component}%
\end{equation}

In this paper, we study a point-like star freely falling along the radial
direction at $\theta_{i}=\pi/2$ and $\varphi_{i}=0$, which emits photons
isotropically in its rest frame. With spherical symmetry, we can confine
ourselves to emissions on the equatorial plane. The geodesics on the
equatorial plane are described by the Lagrangian
\begin{equation}
\mathcal{L}=-\frac{1}{2}\left[  f_{i}(r_{i})\dot{t}_{i}^{2}+\frac{1}%
{f_{i}(r_{i})}\dot{r}_{i}^{2}+r_{i}^{2}\dot{\varphi_{i}}^{2}\right]  ,
\label{eq:Lagrangian}%
\end{equation}
where dots stand for derivative with respect to an affine parameter $\tau$.
Since the Lagrangian $\mathcal{L}$ does not depend on coordinates $t_{i}$ and
$\varphi_{i}$, the geodesics can be characterized by their conserved energy
$E_{i}$ and angular momentum $l_{i}$ in $\mathcal{M}_{i}$,
\begin{equation}
E_{i}=-p_{t_{i}}=f_{i}(r_{i})\dot{t_{i}}\text{, }l_{i}=p_{\varphi_{i}}%
=r_{i}^{2}\dot{\varphi_{i}}. \label{eq:canonical momentums}%
\end{equation}
Note that, according to eqn. $\left(  \ref{eq:connect-component}\right)  $,
one has $E_{1}=E_{2}/Z$ and $l_{1}=l_{2}$.

The Lagrangian of the freely-falling star obeys the constancy $\mathcal{L}%
=-1/2$ when the affine parameter $\tau$ is chosen as the proper time. Since
the star falls radially, its angular momentum $l_{i}=0$. Due to the
traversability of the wormhole, we consider two scenarios with distinct
trajectories of the star. In the scenario I, the star with energy $E_{1}=1/Z$
($E_{2}=1$) has a nonzero initial velocity at spatial infinity of
$\mathcal{M}_{1}$. So, the star can pass through the throat and travel towards
spatial infinity of $\mathcal{M}_{2}$. With the relation $\left(
\ref{eq:canonical momentums}\right)  $, the four-velocities of the star in
$\mathcal{M}_{1}$ and $\mathcal{M}_{2}$ are given by
\begin{align}
v_{e}^{\mu_{1}}(r_{1})  &  =\left(  \frac{1}{1-2r_{1}^{-1}}\sqrt{\frac
{R-2}{R-2k}},-\sqrt{\frac{2k-2}{R-2k}+\frac{2}{r_{1}}},0,0\right)
,\nonumber\\
v_{e}^{\mu_{2}}(r_{2})  &  =\left(  \frac{1}{1-2kr_{2}^{-1}},\sqrt{\frac
{2k}{r_{2}}},0,0\right)  . \label{eq:v SI}%
\end{align}
In the scenario II, the star with energy $E_{1}=1$ is initially at rest at
spatial infinity of $\mathcal{M}_{1}$. At first, the star falls freely in
$\mathcal{M}_{1}$, passes through the throat and reaches a turning point in
$\mathcal{M}_{2}$. Then, it moves towards the throat in $\mathcal{M}_{2}$,
returns to $\mathcal{M}_{1}$ and comes to rest at spatial infinity of
$\mathcal{M}_{1}$. Similarly, the four-velocities of the star in
$\mathcal{M}_{1}$ and $\mathcal{M}_{2}$ are
\begin{align}
v_{e}^{\mu_{1}}(r_{1})  &  =\left(  \frac{1}{1-2r_{1}^{-1}},\mp\sqrt{\frac
{2M}{r_{1}}},0,0\right)  ,\nonumber\\
v_{e}^{\mu_{2}}(r_{2})  &  =\left(  \frac{1}{1-2kr_{2}^{-1}}\sqrt{\frac
{R-2k}{R-2}},\pm\sqrt{\frac{-2k+2}{R-2}+\frac{2k}{r_{2}}},0,0\right)  ,
\label{eq:v SII}%
\end{align}
where plus and minus signs represent outward and inward moving, respectively.

Moreover, null geodesics on the equatorial plane are also governed by the
Lagrangian $\left(  \ref{eq:Lagrangian}\right)  $ with $\mathcal{L}=0$, which
rewrites the radial component of the null geodesic equations as
\begin{equation}
\frac{\dot{r_{i}}^{2}}{L_{i}^{2}}=\frac{1}{b_{i}^{2}}-V_{i\text{,eff}}\left(
r_{i}\right)  , \label{eq:r-geo}%
\end{equation}
where $b_{i}\equiv l_{i}/E_{i}$ is the impact parameter, and $V_{i\text{,eff}%
}\left(  r_{i}\right)  =f_{i}(r_{i})r_{i}^{-2}$ is the effective potential.
Note that the impact parameters of a null geodesic in $\mathcal{M}_{1}$ and
$\mathcal{M}_{2}$, namely $b_{1}$ and $b_{2}$, are related by $b_{1}=Zb_{2}$.
A photon sphere in $\mathcal{M}_{i}$ is constituted of unstable circular null
geodesics, whose radius $r_{i}^{\text{ph}}$ is determined by
\begin{equation}
V_{i\text{,eff}}(r_{i}^{\text{ph}})=\frac{1}{(b_{i}^{\text{ph}})^{2}}\text{,
}V_{i\text{,eff}}^{^{\prime}}(r_{i}^{\text{ph}})=0\text{, }V_{i\text{,eff}%
}^{^{\prime\prime}}(r_{i}^{\text{ph}})<0,
\end{equation}
where $b_{i}^{\text{ph}}$ is the corresponding impact parameter. Photons with
$b_{i}\approx b_{i}^{\text{ph}}$ are temporarily trapped at the photon sphere
and can determine late-time observational appearances of the wormhole. If the
throat radius satisfies $\max\{2,2k\}<R<\min\{3,3k\}$, the asymmetric
thin-shell wormhole can be free of the event horizon and possess two photon
spheres, which are located at $r_{1}^{\text{ph}}=3$ and $r_{2}^{\text{ph}}=3k$
in $\mathcal{M}_{1}$ and $\mathcal{M}_{2}$, respectively. In this paper, we
consider the asymmetric thin-shell wormhole with $k=1.2$ and $R=2.6$, whose
observational appearance of an accretion disk has been discussed in
\cite{Peng:2021osd}.

\begin{table}[ptb]
 \renewcommand\arraystretch{1.3}
\begin{tabular}{>{\raggedright}m{2cm}>{\centering}p{1.5cm}>{\centering}p{6cm}>{\centering}p{6cm}}
	\hline 
	\multicolumn{2}{c}{} & Inward & Outward\tabularnewline
	\hline 
	\multirow{2}{2cm}{Scenario I} & $\mathcal{M}_{1}$ & $\sqrt{\frac{R-2}{R-2k}}-\cos(\alpha)\sqrt{\frac{2k-2}{R-2k}+\frac{2}{r_{e}}}$ & /\tabularnewline
	& $\mathcal{M}_{2}$ & / & $\sqrt{\frac{R-2}{R-2k}}+\cos(\alpha)\sqrt{\frac{R-2}{R-2k}}\sqrt{\frac{2k}{r_{e}}}$\tabularnewline
	\hline 
	\multirow{2}{2cm}{Scenario II} & $\mathcal{M}_{1}$ & $1-\cos(\alpha)\sqrt{\frac{2}{r_{e}}}$ & $1+\cos(\alpha)\sqrt{\frac{2}{r_{e}}}$\tabularnewline
	& $\mathcal{M}_{2}$ & $1-\cos(\alpha)\sqrt{\frac{R-2}{R-2k}}\sqrt{\frac{-2k+2}{R-2}+\frac{2k}{r_{e}}}$ & $1+\cos(\alpha)\sqrt{\frac{R-2}{R-2k}}\sqrt{\frac{-2k+2}{R-2}+\frac{2k}{r_{e}}}$\tabularnewline
	\hline 
\end{tabular}
\caption{The normalized
frequency $\omega_{o}/\omega_{e}$ as a function of the star position $r_{e}$
and the emission angle $\alpha$ in the scenarios I and II. Inward and outward
correspond to travelling towards and away from the throat, respectively.}%
\label{Table: wo/we}%
\end{table}

We assume that the emitted photons are collected by distant observers
distributed on a celestial sphere located at $r_{1}=r_{o}$ in $\mathcal{M}%
_{1}$. To trace light rays emitting from the star to a distant observer, one
needs to supply initial conditions. For a photon of four-momentum $p_{\mu_{i}%
}$, the momentum measured in the rest frame of the star with four-velocity
$v_{e}^{\mu_{i}}$ at $r_{i}=r_{e}$ is
\begin{align}
p^{\hat{t}}  &  =-v_{e}^{t_{i}}(r_{e})p_{t_{i}}-v_{e}^{r_{i}}(r_{e})p_{r_{i}%
},\nonumber\\
p^{\hat{r}}  &  =-\sqrt{\left[  v_{e}^{t_{i}}(r_{e})\right]  ^{2}-f_{i}%
^{-1}(r_{e})}p_{t_{i}}\pm\sqrt{\left[  v_{e}^{r_{i}}(r_{e})\right]  ^{2}%
+f_{i}(r_{e})}p_{r_{i}},\label{eq:p-rest}\\
p^{\hat{\theta}}  &  =0\text{, }p^{\hat{\varphi}}=\frac{p_{\varphi_{i}}}%
{r_{e}},\nonumber
\end{align}
where plus and minus signs correspond to negative and positive $v_{e}^{r_{i}}%
$, respectively. The emission angle $\alpha$ is defined as
\begin{equation}
\cos\alpha=\frac{p^{\hat{r}}}{p^{\hat{t}}}, \label{eq:beta}%
\end{equation}
which is the angle between the propagation direction of the photon and the
radial direction in the rest frame of the star. In the rest frame, the photon
is emitted with proper frequency $\omega_{e}=-\left(  v_{e}^{\mu_{i}}%
p_{\mu_{i}}\right)  _{e}=p^{\hat{t}}$. For a distant static observer with
four-velocity $v_{o}^{\mu_{1}}=\left(  1,0,0,0\right)  $, the photon is
observed with frequency $\omega_{o}=-\left(  v_{o}^{\mu_{1}}p_{\mu_{1}%
}\right)  _{o}=p^{t_{1}}$. With eqns. $\left(  \ref{eq:connect-component}%
\right)  $, $\left(  \ref{eq:p-rest}\right)  $ and $\left(  \ref{eq:beta}%
\right)  $, we express the normalized frequency $\omega_{o}/\omega_{e}$ as a
function of the star position $r_{e}$ and the emission angle $\alpha$ for two
scenarios in Table. \ref{Table: wo/we}. Furthermore, the luminosity of photons
is given by $L_{k}=d\mathcal{E}_{k}/d\tau_{k}$, where $\mathcal{E}_{k}$ is the
total energy, $\tau_{k}$ is the proper time, and $k=e$ and $o$ denote
quantities corresponding to the emitter and the observer, respectively.
Similar to the normalized frequency, one can define the normalized luminosity
\begin{equation}
\frac{L_{o}}{L_{e}}=\frac{d\mathcal{E}_{o}/d\tau_{o}}{d\mathcal{E}_{e}%
/d\tau_{e}}\approx\frac{\omega_{o}dn_{o}}{\omega_{e}dn_{e}}\left(
\frac{dt_{o}}{d\tau_{e}}\right)  ^{-1},
\end{equation}
where $n_{o}$ and $n_{e}$ are the observed and emitted photon numbers,
respectively, and we replaced $d\tau_{o}$ by $dt_{o}$ since they are almost
the same for distant observers.

\section{Numerical Results}

\label{sec:Numerical results}

In this section, we numerically study observational appearances of a star
freely falling radially in the asymmetric thin-shell wormhole in the scenarios
I and II. During the free fall of the star, photons are emitted isotropically
in the rest frame of the star. Specifically, we assume that the star starts
emitting photons at $t_{1}=t_{2}=0$ and $r_{1}=30.65$ in $\mathcal{M}_{1}$,
and emits $3200$ photons, which are uniformly distributed in the emission
angle $\alpha$, every proper time interval $\delta\tau_{e}=0.002$. It is worth
emphasizing that observational appearances of the freely-falling star,
especially late-time appearances, are rather insensitive to the initial
position where the star starts emitting. Here, for better comparison with the
Schwarzschild black hole case, we simply choose the initial position as
$r_{1}=30.65$, which is in agreement with that of \cite{Cardoso:2021sip}.

Here, observational appearances of the star are studied for two kinds of
observers in $\mathcal{M}_{1}$. The first kind is observers distributed on a
celestial sphere at the radius $r_{o}=100$, which refers to collecting photons
in the whole sky at fixed radial coordinate $r_{o}=100$ in $\mathcal{M}_{1}$.
The measurement by the observers on the celestial sphere would give the
frequency distribution and the total luminosity of photons that reach the
celestial sphere. The second kind is a specific observer, who is located at
$\varphi_{o}=0$ on the equator of the celestial sphere. Among all photons
collected on the celestial sphere, we select photons with $\cos\varphi>0.99$
to mimic photons detected by the specific observer. To calculate observed
luminosities, the collected photons are grouped into packets of 50 (i.e.,
$dn_{o}=50$) according to their arrival time.

\begin{figure}[ptb]
\includegraphics[width=0.65\textwidth]{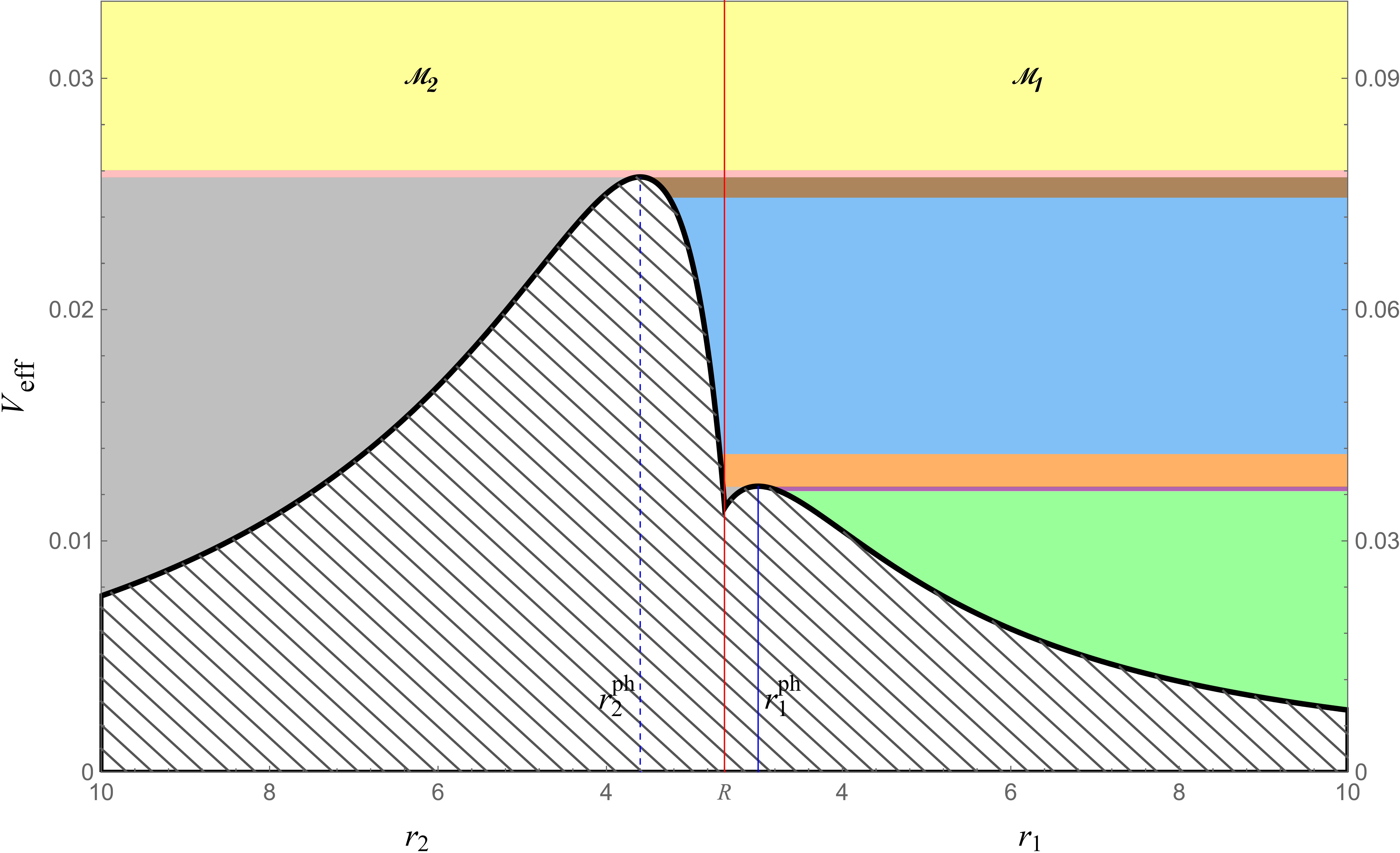}\caption{The effective potential of
null geodesics in the asymmetric thin-shell wormhole with $k=1.2$ and $R=2.6$.
The potential has two peaks at $r_{1}^{\text{ph}}=3$ (solid vertical blue
line) and $r_{2}^{\text{ph}}=3.6$ (dashed vertical blue line), corresponding to
a photon sphere with $b_{1}^{\text{ph}}=3\sqrt{3}$ in $\mathcal{M}_{1}$ and
another one with $b_{2}^{\text{ph}}=3.6\sqrt{3}$ in $\mathcal{M}_{2}$,
respectively. The vertical red line denotes the throat at $r_{1}=r_{2}=R$.
Photons emitted in the pink, brown, orange and purple regions have impact
parameters close to the impact parameters of the photon spheres, and hence can
be temporarily trapped around the photon spheres. In particular, when photons
are emitted towards the throat at $r_{2}>r_{2}^{\text{ph}}$ in the pink region
or at $r_{1}>r_{1}^{\text{ph}}$ in the brown, orange and purple regions, they
usually orbit the wormhole with $\Delta\varphi\geq2\pi$.}%
\label{Fig: Veff-WH}%
\end{figure}

\begin{figure}[ptb]
\includegraphics[width=0.9\textwidth]{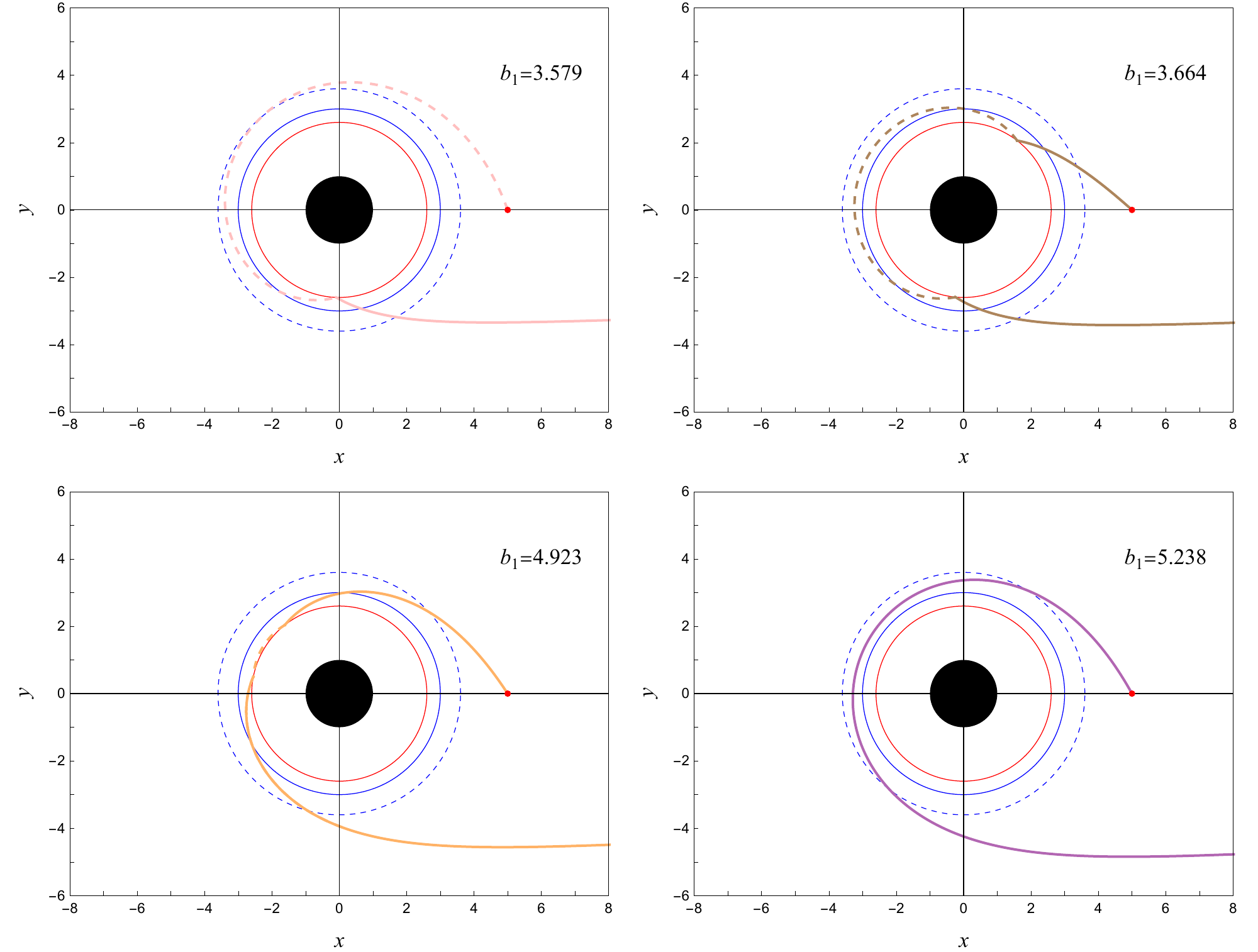}\caption{Photon trajectories in the
asymmetric thin-shell wormhole with $k=1.2$ and $R=2.6$. The red points and
circles denote the star and the throat, respectively. The blue solid and
dashed circles represent the photon spheres in $\mathcal{M}_{1}$ and
$\mathcal{M}_{2}$, respectively. The upper-left panel shows a photon emitted
at $r_{e}=5$ in $\mathcal{M}_{2}$ with $b_{1}=3.579$, and the light ray has
$\Delta\varphi=2\pi$. Other panels show photons emitted at $r_{e}=5$ in
$\mathcal{M}_{1}$ with $b_{1}=3.664$, $4.923$ and $5.238$, and the light rays
all have $\Delta\varphi=2\pi$. The solid and dashed segments of the light rays
correspond to the segments in $\mathcal{M}_{1}$ and $\mathcal{M}_{2}$,
respectively.}%
\label{Fig: tr-WH}%
\end{figure}

As shown in FIG. \ref{Fig: Veff-WH}, the asymmetric thin-shell wormhole with
$k=1.2$ and $R=2.6$ has a double-peak effective potential, corresponding to
one photon sphere in $\mathcal{M}_{1}$ and one in $\mathcal{M}_{2}$.
Specifically, the photon sphere in $\mathcal{M}_{1}$ is located at
$r_{1}^{\text{ph}}=3$ with the critical impact parameter $b_{1}^{\text{ph}%
}=3\sqrt{3}$, and that in $\mathcal{M}_{2}$ is located at $r_{2}^{\text{ph}%
}=3.6$ with the critical impact parameter $b_{2}^{\text{ph}}=3.6\sqrt{3}$. To
discuss how photons with different impact parameters contribute to the
observations of the star, we classify received photons into seven categories
according to their impact parameter $b_{1}$ in $\mathcal{M}_{1}$,

\begin{itemize}
\item $b_{1}<3.579$. Yellow region in FIG. \ref{Fig: Veff-WH} and yellow dots
in FIGs. \ref{Fig: sky-WHSI}, \ref{Fig: phi0-WHSI}, \ref{Fig: sky-WHSII} and
\ref{Fig: phi0-WHSII}.

\item $3.579\leq b_{1}<Zb_{2}^{\text{ph}}$. Pink region in FIG.
\ref{Fig: Veff-WH} and pink dots in FIGs. \ref{Fig: sky-WHSI},
\ref{Fig: phi0-WHSI}, \ref{Fig: sky-WHSII} and \ref{Fig: phi0-WHSII}. In this
category, photons emitted inward outside the photon sphere in $\mathcal{M}%
_{2}$ can circle around the photon sphere more than once before reaching a
distant observer in $\mathcal{M}_{1}$. For example, a light ray with
$b_{1}=3.579$, which has $\Delta\varphi=2\pi$\footnote{\label{ft:2} Since
$\varphi_{1}=\varphi_{2}$ at the throat, the subscript of $\varphi$ is omitted
for simplicity.}, is displayed in the upper-left panel of FIG.
\ref{Fig: tr-WH}.

\item $Zb_{2}^{\text{ph}}<b_{1}\leq3.664$. Brown region in FIG.
\ref{Fig: Veff-WH} and brown dots in FIGs. \ref{Fig: sky-WHSI},
\ref{Fig: phi0-WHSI}, \ref{Fig: sky-WHSII} and \ref{Fig: phi0-WHSII}. In this
category, photons emitted inward would circle around the photon sphere in
$\mathcal{M}_{2}$ roughly with $\Delta\varphi\geq2\pi$ before escaping to the
celestial sphere in $\mathcal{M}_{1}$. For example, a light ray with
$b_{1}=3.664$, which has $\Delta\varphi=2\pi$, is displayed in the upper-right
panel of FIG. \ref{Fig: tr-WH}.

\item $3.664<b_{1}\leq4.923$. Blue region in FIG. \ref{Fig: Veff-WH} and blue
dots in FIGs. \ref{Fig: sky-WHSI}, \ref{Fig: phi0-WHSI}, \ref{Fig: sky-WHSII}
and \ref{Fig: phi0-WHSII}.

\item $4.923<b_{1}<b_{1}^{\text{ph}}$. Orange region in FIG.
\ref{Fig: Veff-WH} and orange dots in FIGs. \ref{Fig: sky-WHSI},
\ref{Fig: phi0-WHSI}, \ref{Fig: sky-WHSII} and \ref{Fig: phi0-WHSII}. In this
category, if photons are emitted inward outside the photon sphere in
$\mathcal{M}_{1}$, they would linger for some time around the photon sphere by
orbiting it approximately with $\Delta\varphi\geq2\pi$. For example, a light
ray with $b_{1}=4.923$, which has $\Delta\varphi=2\pi$, is displayed in the
lower-left panel of FIG. \ref{Fig: tr-WH}.

\item $b_{1}^{\text{ph}}<b_{1}\leq5.238$. Purple region in FIG.
\ref{Fig: Veff-WH} and purple dots in FIGs. \ref{Fig: sky-WHSI},
\ref{Fig: phi0-WHSI}, \ref{Fig: sky-WHSII} and \ref{Fig: phi0-WHSII}. In this
category, photons emitted inward outside the photon sphere in $\mathcal{M}%
_{1}$ usually circle around the photon sphere more than once. For example, a
light ray with $b_{1}=5.238$, which has $\Delta\varphi=2\pi$, is displayed in
the lower-right panel of FIG. \ref{Fig: tr-WH}.

\item $b_{1}>5.238$. Green region in FIG. \ref{Fig: Veff-WH} and green dots in
FIGs. \ref{Fig: sky-WHSI}, \ref{Fig: phi0-WHSI}, \ref{Fig: sky-WHSII} and
\ref{Fig: phi0-WHSII}.
\end{itemize}

In short, we use the orbit number of light rays emitted at $r_{1}=5$ in
$\mathcal{M}_{1}$ or $r_{2}=5$ in $\mathcal{M}_{2}$ to determine the threshold
impact parameters separating the seven categories. To sum up, light rays
emitted inward at $r_{2}=5$ in the yellow/pink category would circle around
the wormhole less/more than once before being received; light rays emitted
inward at $r_{1}=5$ would circle around the wormhole less than once before
being received in the blue and green categories, or more than once in the
brown, orange and purple categories. Note that the orbit number of light rays
with a given impact parameter depends slightly on the emitting position. So,
light rays connecting the star and the observers circle around the wormhole
approximately more than once in the pink, brown, orange and purple categories,
and less than once in the yellow, blue and green categories. In other words,
photons in the pink, brown, orange and purple categories can be temporarily
trapped near the photon spheres.

\subsection{Scenario I}

\label{sub-sec:Scenario I}

\begin{figure}[ptb]
\includegraphics[width=0.55\textwidth]{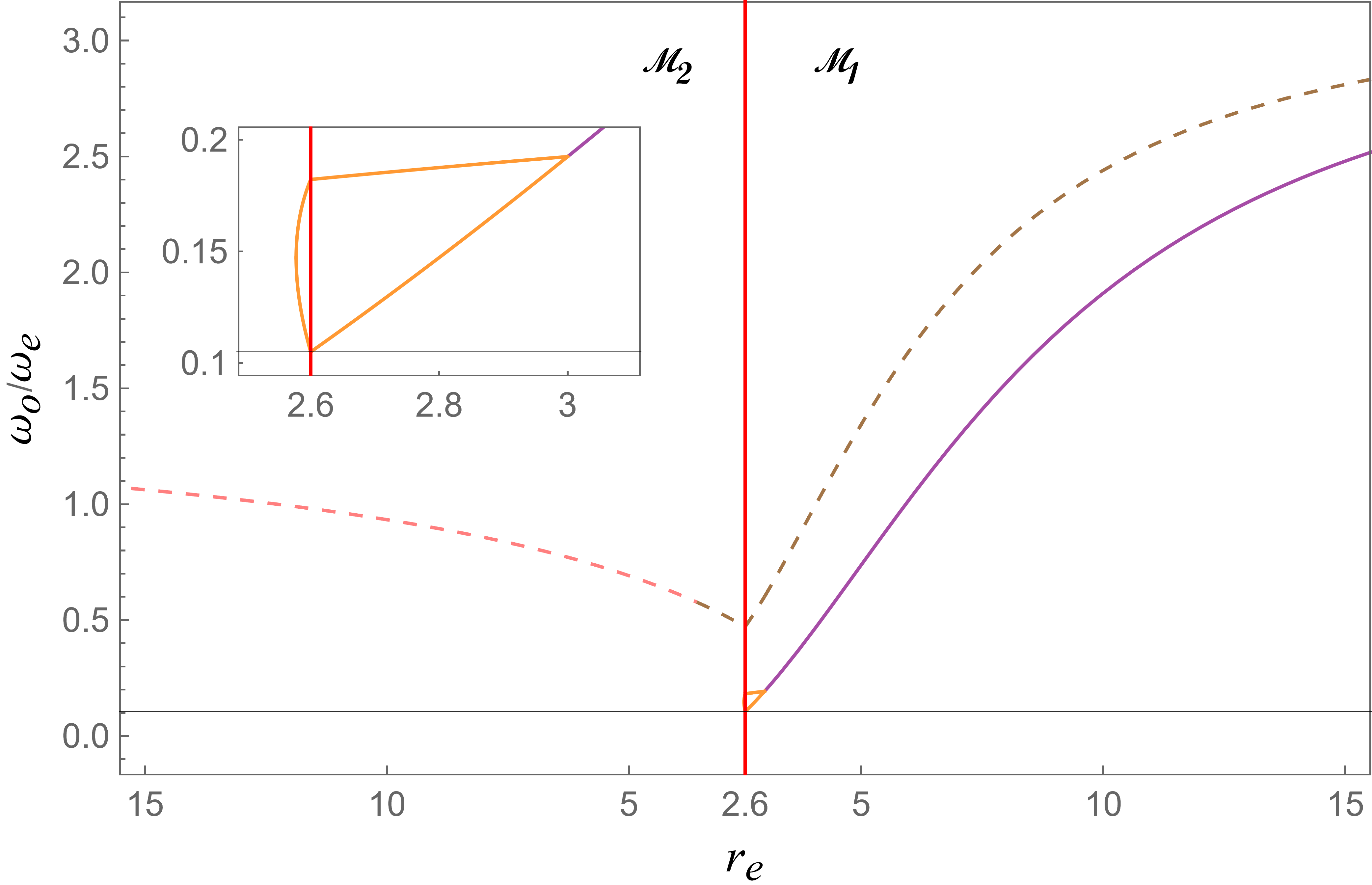}\caption{The normalized frequency
$\omega_{o}/\omega_{e}$ as a function of the emitted position $r_{e}$ for
photons in the scenario I, whose impact parameter is very close to these of
the photon spheres in $\mathcal{M}_{1}$ (solid lines) and $\mathcal{M}_{2}$
(dashed lines). The observers are distributed on the celestial sphere at
$r_{o}=100$ in $\mathcal{M}_{1}$. For a large $r_{e}$ in $\mathcal{M}_{1}$,
inward-emitted and near-critical photons can be blueshifted since the Doppler
effect dominates over the gravitational redshift. Due to the relation $\left(
\ref{eq:connect-component}\right)  $ at the throat, near-critical photons can
also be blueshifted when $r_{e}$ is large in $\mathcal{M}_{2}$. Photons
emitted inward and outward between the two photon spheres can both reach a
distant observer after orbiting the photon sphere in $\mathcal{M}_{1}$, which
gives two branches of the orange line in the inset. Moreover, the normalized
frequency reaches the minimum at the throat, which is located at $r_{e}=2.6$.}%
\label{Fig: bc-SI}%
\end{figure}

In the scenario I, the star with energy $E_{1}=1/Z=\sqrt{3}$ would travel
through the throat and move towards spatial infinity of $\mathcal{M}_{2}$. For
near-critical photons emitted with the impact parameter very close to those of
the photon spheres in $\mathcal{M}_{1}$ (i.e., $b_{1}\simeq b_{1}^{\text{ph}}%
$) and $\mathcal{M}_{2}$ (i.e., $b_{2}\simeq b_{2}^{\text{ph}}$), their
normalized frequencies $\omega_{o}/\omega_{e}$ measured by observers on the
celestial sphere are plotted against the emitted position $r_{e}$ in FIG.
\ref{Fig: bc-SI}. The colors of the lines in FIG. \ref{Fig: bc-SI} match those
of the corresponding emitted regions in FIG. \ref{Fig: Veff-WH}. Moreover,
photons with $b_{1}\simeq b_{1}^{\text{ph}}$ and $b_{2}\simeq b_{2}%
^{\text{ph}}$ are denoted by solid and dashed lines, respectively. It is worth
emphasizing that the observed frequency of a photon is determined by the
gravitational redshift and the Doppler effect, which are controlled by the
position and the velocity of the photon when it is emitted, respectively.

For photons of $b_{2}\simeq b_{2}^{\text{ph}}$, the normalized frequency can
noticeably exceed $1$ at a large $r_{e}$ in $\mathcal{M}_{1}$ since the
Doppler effect plays a more important role than the gravitational redshift. As
the star falls towards the throat, the normalized frequency decreases due to
stronger gravitational redshift, and blueshift becomes redshift at
$r_{e}=4.063$ in $\mathcal{M}_{1}$, where the normalized frequency is 1. When
emitted at the throat, the normalized frequency reaches the minimum. After the
star enters $\mathcal{M}_{2}$, the normalized frequency increases as $r_{e}$
grows, and observed photons become bluershifted when $r_{e}>12.281$. For
photons of $b_{1}\simeq b_{1}^{\text{ph}}$, the behavior of the normalized
frequency is quite similar to those of $b_{2}\simeq b_{2}^{\text{ph}}$ when
they are emitted outside the photon sphere in $\mathcal{M}_{1}$. When the star
emits photons between the two photon spheres, inward-emitted and
outward-emitted photons can both be captured by a distant observer after they
circle around the photon sphere in $\mathcal{M}_{1}$, thus leading to two
branches as shown in the inset. The upper and lower branches correspond to
photons emitted away from and towards the observer, respectively.

\begin{figure}[ptb]
\includegraphics[width=1\textwidth]{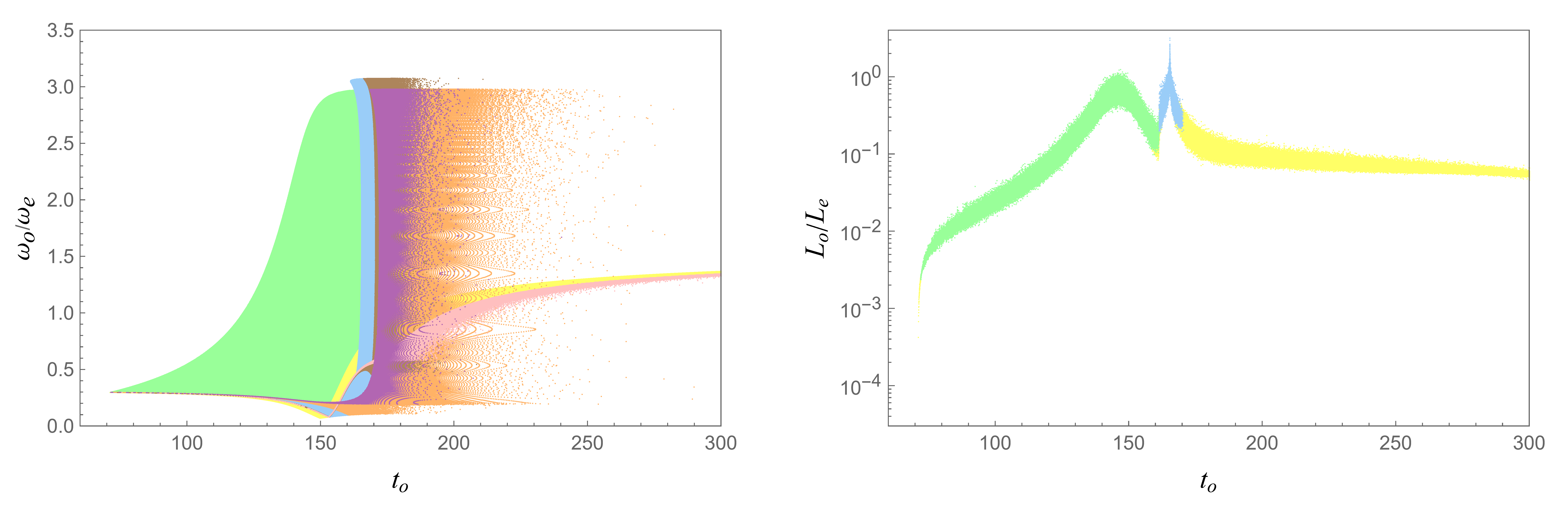}\caption{The normalized frequency
distribution and the total luminosity of the freely-falling star in the
scenario I, measured by observers on a celestial sphere at $r_{o}=100$ in
$\mathcal{M}_{1}$. \textbf{Left:} The observers receive photons with a wide
range of frequencies. At the early stage, photons emitted in the green region
of FIG. \ref{Fig: Veff-WH} give rise to the frequency observation. Afterwards,
photons emitted in the brown, blue, orange and purple regions are observed. In
particular, photons with a near-critical impact parameter produce high
frequency observations. The late-time frequency observation is determined by
photons in the yellow and pink regions, which are emitted at a large $r_{e}$
in $\mathcal{M}_{2}$. \textbf{Right:} The luminosity is calculated by grouping
received photons into packets of 50. An increase of the observed luminosity is
caused by photons emitted inward in the blue region, leading to a peak at
$t_{o}\simeq168$. At late times, the total luminosity gradually decays with
time and is mainly controlled by photons, which are emitted at a large $r_{e}$
in $\mathcal{M}_{2}$ and travel through the throat to reach the observers.}%
\label{Fig: sky-WHSI}%
\end{figure}

In the left panel of FIG. \ref{Fig: sky-WHSI}, we display the normalized
frequency distribution of photons, which are emitted from the freely-falling
star in the scenario I and collected by observers distributed on the celestial
sphere at $r_{o}=100$ in $\mathcal{M}_{1}$. At early times, received photons
are dominated by those emitted in the green region of FIG. \ref{Fig: Veff-WH},
among which inward-emitted photons contribute to the high-frequency
observation. When $t_{o}>160$, photons emitted towards the photon sphere in
$\mathcal{M}_{2}$ in the blue and brown regions start reaching the observers
after orbiting around the photon sphere. Subsequently, the observers receive
photons emitted towards the photon sphere in $\mathcal{M}_{1}$ in the purple
and orange regions. Since time moves faster in $\mathcal{M}_{2}$ roughly by a
factor of $1/Z=\sqrt{3}$ relative to in $\mathcal{M}_{1}$, photons circling
around the photon sphere in $\mathcal{M}_{2}$ arrive earlier. Moreover, the
maximum frequency of photons emitted in the blue and brown regions is higher
than that of photons emitted in the green, purple and orange regions. This is
expected from FIG. \ref{Fig: bc-SI}, which shows that near-critical photons
with $b_{2}\simeq b_{2}^{\text{ph}}$ have higher normalized frequency than
these with $b_{1}\simeq b_{1}^{\text{ph}}$. Afterwards, the frequency
observations are dominated by photons emitted in the orange region, which are
trapped at the photon sphere in $\mathcal{M}_{1}$ for a longer time. At late
times, the observers mostly receive photons in the yellow and pink regions,
which are emitted towards the throat in $\mathcal{M}_{2}$ with a small impact parameter.

The normalized total luminosity of the freely-falling star is displayed in the
right panel of FIG. \ref{Fig: sky-WHSI}, where a dot corresponds to a packet
of 50 photons, and the color of the dot is that having most photons in the
packet. The luminosity gradually increases until reaching a peak around
$t_{o}\simeq145$, and is dominated by photons emitted in the green region
roughly before $t_{o}=150$, which is in agreement with the frequency
observation. After $t_{o}\simeq160$, photons emitted in the blue region give
rise to a noticeable increase of the total luminosity. As the star moves
towards spatial infinity of $\mathcal{M}_{2}$, emitted photons can still
propagate to the observers in $\mathcal{M}_{1}$ through the throat, and a
slight decrease of the total luminosity is displayed at late times.
Interestingly, this late-time observation is strikingly different from the
black hole case, where the total luminosity has been found to decay
exponentially at late times \cite{Cardoso:2021sip,Chen:2022qrw}.

\begin{figure}[ptb]
\includegraphics[width=1\textwidth]{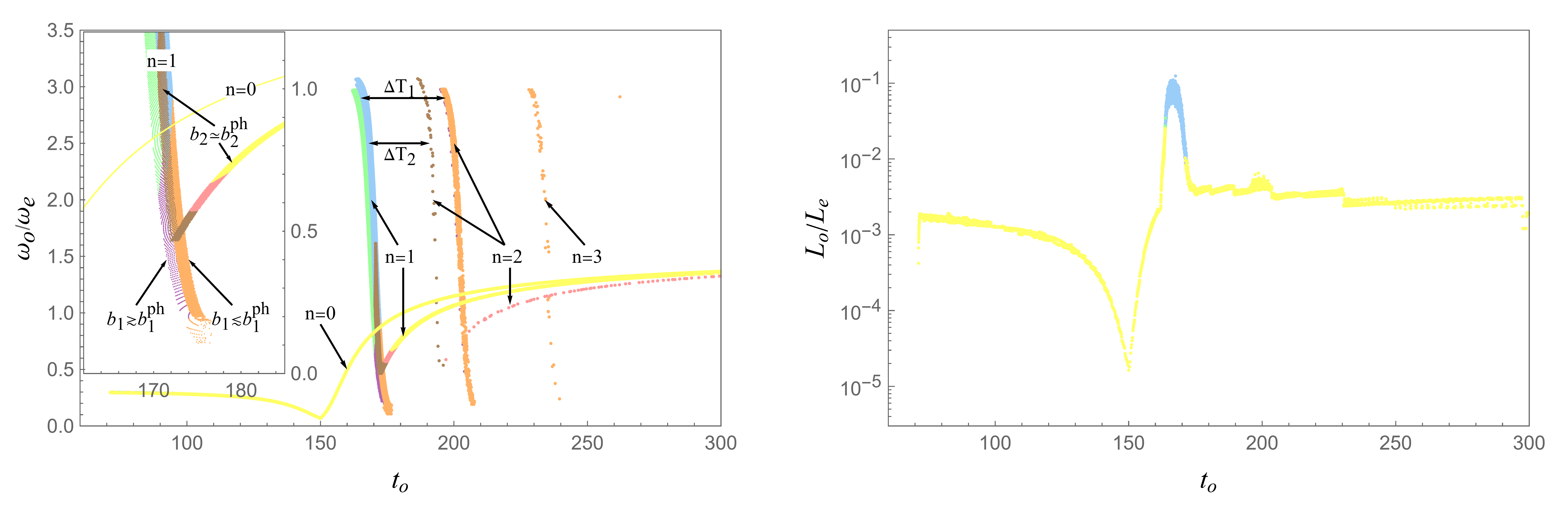}\caption{The normalized
frequency and the luminosity of the freely-falling star in the scenario I,
measured by a distant observer at $r_{o}=100$, $\theta_{o}=\pi/2$ and
$\phi_{o}=0$ in $\mathcal{M}_{1}$. The colored dots denote photons emitted in
the regions with the same color in FIG. \ref{Fig: Veff-WH}. \textbf{Left:}
Received photons form several frequency lines indexed by the orbiting number
$n$. The inset displays three frequency lines caused by $n=1$ photons with
$b_{1}\gtrsim b_{1}^{\text{ph}}$, $b_{1}\lesssim b_{1}^{\text{ph}}$ and
$b_{2}\simeq b_{2}^{\text{ph}}$. The time delay between the adjacent $n\geq1$
lines formed by photons orbiting around the photon sphere in $\mathcal{M}_{1}$
and $\mathcal{M}_{2}$ is roughly the period of circular null geodesics at the
photon sphere, i.e., $\Delta T_{1}\simeq2\pi b_{1}^{\text{ph}}\simeq33$ and
$\Delta T_{2}\simeq2\pi Zb_{2}^{\text{ph}}\simeq23$, respectively.
\textbf{Right: }At early times, the luminosity is dominated by photons with a
small impact parameter, and decreases first and then increases after the star
goes through the throat. Subsequently, blueshifted $n=1$ photons start to
reach the observer and become the most dominant contribution, which produces a
luminous flash at $t_{o}\simeq170$. Later, the luminosity is mainly
contributed by the $n=0$ photons emitted in the yellow region of
$\mathcal{M}_{2}$ and almost declines gradually at late times. In addition, a
faint flash, which results from the $n=2$ photons emitted in the orange
region, is observed at $t_{o}\simeq200$.}%
\label{Fig: phi0-WHSI}%
\end{figure}

For a specific observer located at $\varphi_{o}=0$ and $\theta_{o}=\pi/2$ on
the celestial sphere at $r_{o}=100$ in $\mathcal{M}_{1}$, the angular
coordinate change $\Delta\varphi$ of light rays connecting the star with the
observer is
\begin{equation}
\Delta\varphi=2n\pi\text{,}%
\end{equation}
where $n=0,1,2\cdots$ is the number of orbits that the light rays complete
around the wormhole. To simulate observational appearances of the star seen by
the observer, we select photons with $\cos\varphi>0.99$ from all
photons received on the celestial sphere. The frequency observation is
presented in the left panel of FIG. \ref{Fig: phi0-WHSI}, which shows a
discrete spectrum separated by the received time. The yellow line is formed by
photons with $n=0$, which radially propagate to the observer. At early times,
the observed frequency of the $n=0$ photons decreases with the received time
as the star falls towards the throat. After the star passes through the
throat, the observed frequency of the $n=0$ photons increases since the
gravitational redshift becomes weaker as the star moves further away from the
throat, which results in the dip at $t_{o}\simeq150$. Owing to the existence
of two photon spheres, the $n=1$ photons with impact parameters $b_{1}\gtrsim
b_{1}^{\text{ph}}$, $b_{1}\lesssim b_{1}^{\text{ph}}$ and $b_{2}\simeq
b_{2}^{\text{ph}}$ can form three frequency lines, which are highlighted in
the inset of FIG. \ref{Fig: phi0-WHSI}. As the star falls towards the throat,
the three frequency lines decrease rapidly due to strong gravitational
redshift near the throat. After the star passes through the throat, the
frequency line with $b_{2}\simeq b_{2}^{\text{ph}}$ gradually increases. For
$n=2$, the frequency lines with $b_{1}\gtrsim b_{1}^{\text{ph}}$ and
$b_{1}\lesssim b_{1}^{\text{ph}}$ move closer and are hardly distinguishable
from each other. On the other hand, the frequency line with $b_{2}\simeq
b_{2}^{\text{ph}}$ becomes more separate from them since photons spend more
time orbiting around the photon sphere in $\mathcal{M}_{1}$. Indeed, it takes
$\Delta T_{1}\simeq2\pi b_{1}^{\text{ph}}\simeq33$ to orbit around the photon
sphere in $\mathcal{M}_{1}$ one time, and $\Delta T_{2}\simeq2\pi
Zb_{2}^{\text{ph}}\simeq23$ to orbit around that in $\mathcal{M}_{2}%
$\footnote{\label{ft:3} Eqn. $\left(  \ref{eq:canonical momentums}\right)  $
leads to $dt/d\phi|_{r^{\text{ph}}}=b^{-1}V_{\text{eff}}^{-1}(r^{\text{ph}%
})=b^{\text{ph}}$, which gives $\Delta T\simeq2\pi b^{\text{ph}}$.}.
Therefore, for $b_{1}\gtrsim b_{1}^{\text{ph}}$ and $b_{1}\lesssim
b_{1}^{\text{ph}}$ ($b_{2}\simeq b_{2}^{\text{ph}}$), the time delay between
the \thinspace$n=1$ and $2$ frequency lines roughly equals to $\Delta T_{1}$
($\Delta T_{2}$). For $n=3$, because of the finite number of photons in our
numerical simulation, only the frequency line with $b_{2}\gtrsim
b_{2}^{\text{ph}}$ can be found and is shown by orange dots around
$t_{o}\simeq230$.

The left panel of FIG. \ref{Fig: phi0-WHSI} shows the observed normalized
luminosity as a function of the time, which exhibits a decline before the star
reaches the throat. After the star moves through the throat, the luminosity
starts to increase since the frequency of received photons grows, which causes
a dip at $t_{o}\simeq150$. Around $t_{o}\simeq160$, blueshifted photons with
$n=1$ start to play a dominant role, leading to a luminous flash around
$t_{o}\simeq170$. Afterwards, the luminosity is mainly dominated by photons
emitted in the yellow region of $\mathcal{M}_{2}$, and slowly decreases except
a faint flash at $t_{o}\simeq200$ caused by the arrival of $n=2$ photons. The
flashes of photons with $n\geq3$ are much fainter and barely visible in the
background of the dominant photons emitted in the yellow region. In contrast,
for a black hole with two photon spheres, a series of flashes with decreasing
luminosity are observed at late times due to photons orbiting around the hairy
black hole different times \cite{Chen:2022qrw}.

\subsection{Scenario II}

\label{sub-sec:Scenario II}

\begin{figure}[ptb]
\includegraphics[width=0.55\textwidth]{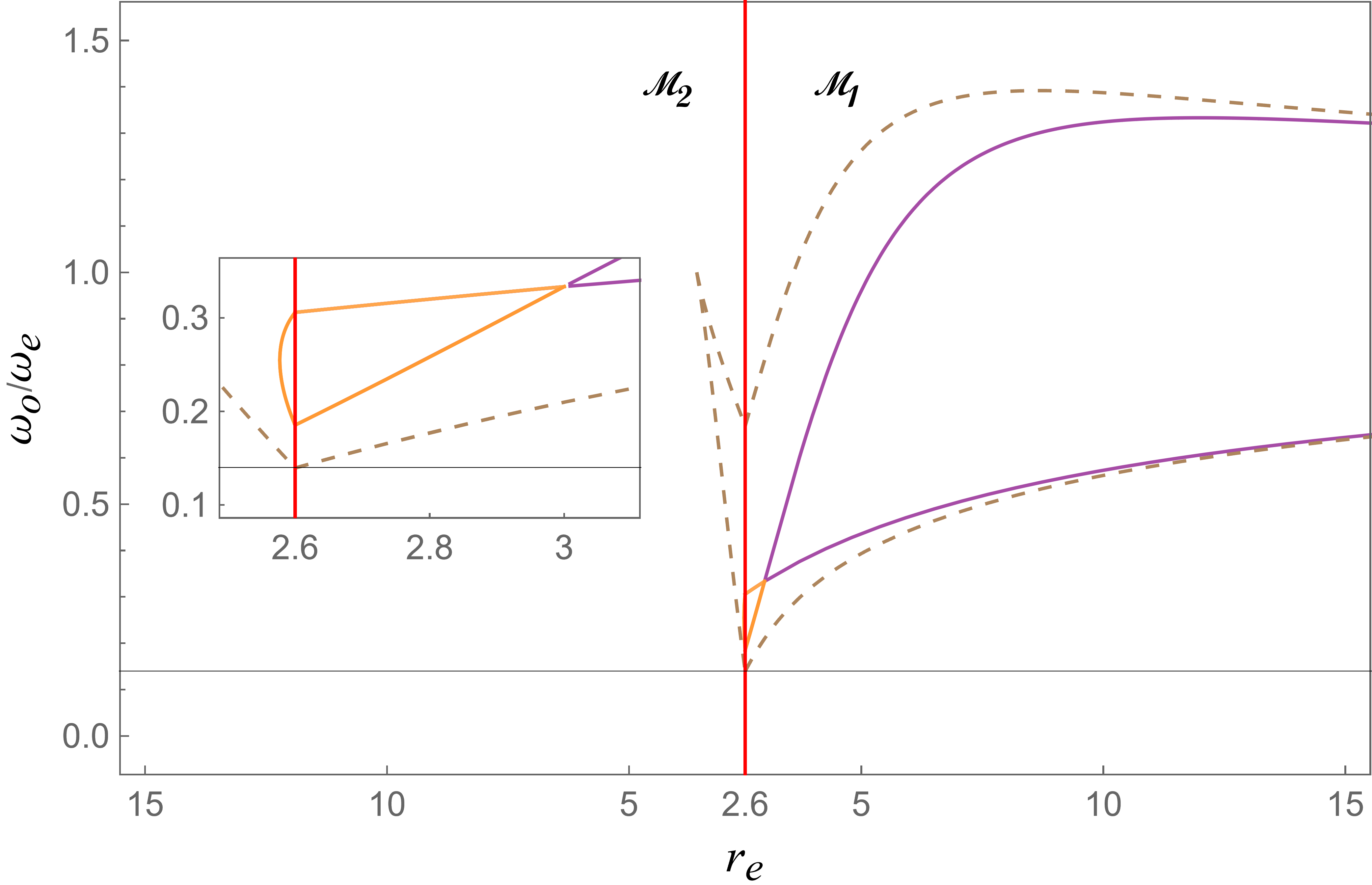}\caption{The normalized frequency
$\omega_{o}/\omega_{e}$ as a function of the emitted position $r_{e}$ for
photons in the scenario II, whose impact parameter $b_{1}$ is very close to
$b_{1}^{\text{ph}}$ (solid lines) or $b_{2}$ is very close to $b_{2}%
^{\text{ph}}$ (dashed lines). The observers are distributed on the celestial
sphere at $r_{o}=100$ in $\mathcal{M}_{1}$. The normalized frequency of
near-critical photons emitted in the purple and brown regions has two
branches. Specifically, the high-frequency (low-frequency) branch corresponds
to photons emitted from the star falling away (towards) from the observer.
Similar to the scenario I, the high-frequency branch can be blueshifted for a
large $r_{e}$ in $\mathcal{M}_{1}$. The normalized frequency reaches the
global minimum $\omega_{o}/\omega_{e}\simeq0.139$ at the throat for the
low-frequency branch.}%
\label{Fig:bc-SII}%
\end{figure}

In the scenario II, the star starts falling from spatial infinity of
$\mathcal{M}_{1}$ and returns to the infinity after going through the throat
twice. Similarly, the normalized frequency $\omega_{o}/\omega_{e}$ for
near-critical photons is plotted in FIG. \ref{Fig:bc-SII}. Specifically, we
focus on photons with $b_{1}\simeq b_{1}^{\text{ph}}$ emitted in the purple
and orange regions and those with $b_{2}\simeq b_{2}^{\text{ph}}$ emitted in
the brown region, which are denoted by solid and dashed lines, respectively.
For photons with $b_{1}\simeq b_{1}^{\text{ph}}$ emitted outside the photon
sphere in $\mathcal{M}_{1}$ (i.e., the purple region) and those with
$b_{2}\simeq b_{2}^{\text{ph}}$, the normalized frequency has high-frequency
and low-frequency branches, corresponding to the star falling away from and
towards the observer, respectively. If photons are emitted inside the photon
sphere in $\mathcal{M}_{1}$ with $b_{1}\simeq b_{1}^{\text{ph}}$, the
high-frequency (low-frequency) branch denotes ingoing and outgoing (outgoing
and ingoing) emissions from the star falling away from and towards the
observer, respectively. \ For the high-frequency branches, strong
gravitational lensing around the photon spheres can cause blueshifts of
near-critical photons emitted inward at a large $r_{e}$ in $\mathcal{M}_{1}$.
In particular, the normalized frequency with $b_{1}\simeq b_{1}^{\text{ph}}$
($b_{2}\simeq b_{2}^{\text{ph}}$) reaches the maximum $\omega_{o}/\omega
_{e}=4/3$ ($\omega_{o}/\omega_{e}=1.392$) at $r_{e}=12$ ($r_{e}=8.679$),
becomes one at $r_{e}=5.196$ ($r_{e}=3.6$), and reaches the minimum
$\omega_{o}/\omega_{e}=0.306$ ($\omega_{o}/\omega_{e}=0.139$) at the throat.
In $\mathcal{M}_{2}$, the normalized frequency with $b_{2}\simeq
b_{2}^{\text{ph}}$ reaches the maximum $\omega_{o}/\omega_{e}=1$ at
$r_{e}=3.6$, where the star returns.

\begin{figure}[ptb]
\includegraphics[width=1\textwidth]{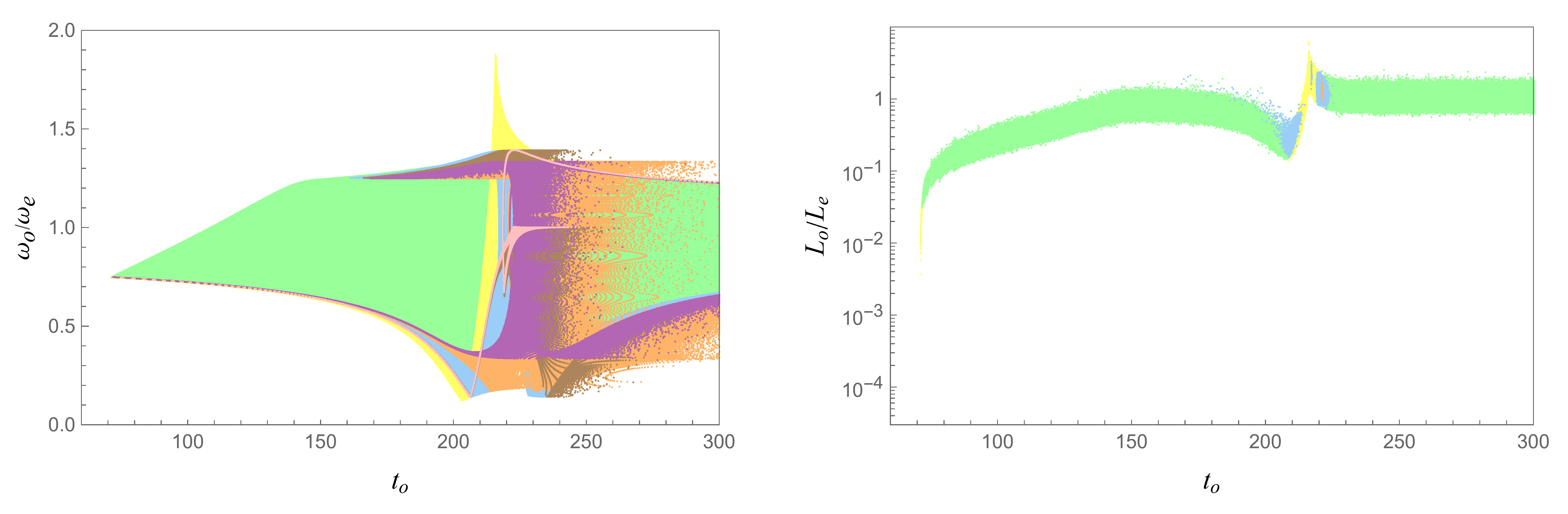}\caption{The normalized
frequency distribution (\textbf{Left}) and the total luminosity
(\textbf{Right}) of the freely-falling star in the scenario II, measured by
observers on the celestial sphere at $r_{o}=100$ in $\mathcal{M}_{1}$. Similar
to the scenario I, photons emitted in the green region of FIG.
\ref{Fig: Veff-WH} dominate the frequency and luminosity observations in the
early stage. After the star enters $\mathcal{M}_{2}$, photons emitted in the
yellow region, which propagates to the observers nearly in the radial
direction, produce frequency and luminosity peaks around $t_{o}\simeq220$.
Later, near-critical photons with a wide range of frequencies are observed. At
late times, the emitted position $r_{e}$ is in $\mathcal{M}_{1}$ and large,
and therefore the observers would collect most of emitted photons, which leads
to a nearly constant total luminosity.}%
\label{Fig: sky-WHSII}%
\end{figure}

The normalized frequency distribution of photons received by observers
distributed on the celestial sphere is presented in the left panel of FIG.
\ref{Fig: sky-WHSII}. When $t_{o}\lesssim200$, a wide range of frequencies is
observed for photons emitted in the green region. After near-critical photons
emitted in the purple, orange, blue and brown regions start arriving at
the observers around $t_{o}\simeq150$, they come to dominate the
high-frequency part of the frequency distribution. This early-stage frequency
distribution bears a resemblance to the Schwarzschild black hole case, in
which a star falls from spatial infinity at rest \cite{Cardoso:2021sip}.
Similar to the scenario I, the maximum frequency of photons emitted inward in
the blue and brown regions is greater than that of photons emitted inward in
the green, purple and orange regions. After the star enters $\mathcal{M}_{2}$,
the observed frequency of photons emitted in the yellow region starts to
increase and reaches a maximum around $t_{o}\simeq220$, which is associated
with the star returning to the throat. Subsequently, photons emitted in the
brown and purple regions are observed to have a wide range of frequencies
after they circle around the photon sphere in $\mathcal{M}_{1}$ and reach the
observers. At late times, the star comes back to $\mathcal{M}_{1}$ and moves
towards the observer, and thus the low-frequency distribution is dominated by
photons emitted towards the throat with $b_{1}\simeq b_{1}^{\text{ph}}$ and
$b_{2}\simeq b_{2}^{\text{ph}}$. On the other hand, photons emitted towards
the observers with a small impact parameter produce the high-frequency observation.

The normalized total luminosity of the freely-falling star in the scenario II
is displayed in the right panel of FIG. \ref{Fig: sky-WHSII}. Before
$t_{o}\simeq200$, the total luminosity behaves similarly to the Schwarzschild
black hole case studied in \cite{Cardoso:2021sip}, which is in consistency
with the frequency observation. Afterwards, the received blueshifted photons
with a small impact parameter dominate the total luminosity, resulting in a
peak at $t_{o}\simeq220$. At late times, the total luminosity is maintained
around one since most emitted photons can be collected by the observers.

\begin{figure}[ptb]
\includegraphics[width=1\textwidth]{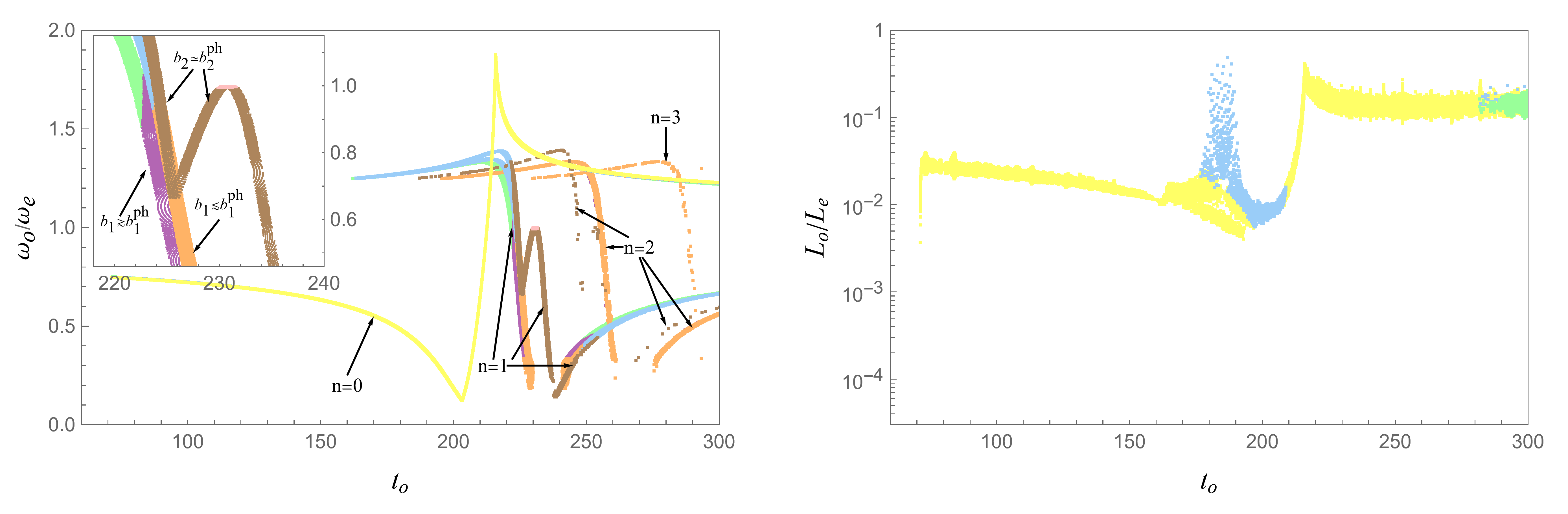}\caption{The normalized
frequency and the luminosity of the freely-falling star in the scenario II,
measured by a distant observer at $r_{o}=100$, $\theta=\pi/2$ and $\phi=0$ in
$\mathcal{M}_{1}$. \textbf{Left:} The yellow line denotes radially emitted
photons with $n=0$ and has a dip (peak) near $t_{o}\simeq200$ ($t_{o}%
\simeq220$), corresponding to emission from the star at the throat. The $n=1$
frequency lines with $b_{1}\gtrsim b_{1}^{\text{ph}}$, $b_{1}\lesssim
b_{1}^{\text{ph}}$ and $b_{2}\simeq b_{2}^{\text{ph}}$ steadily increase to a
peak followed by a sharp decrease when $t_{o}\lesssim230$, and gradually
increase when $t_{o}\gtrsim240$. For $230\lesssim t_{o}\lesssim240$, the $n=1$
frequency line with $b_{2}\simeq b_{2}^{\text{ph}}$ rises to another high
point. \textbf{Right: }Similar to the scenario I, the luminosity is dominated
by $n=0$ photons and gradually decreases before $t_{o}\simeq160$. Later,
blueshifted $n=1$ photons start to reach the observer and then become the most
dominant contribution, which results in a luminosity peak around $t_{o}%
\simeq180$. Afterwards, due to the increasing frequency of $n=0$ photons
emitted in $\mathcal{M}_{2}$, the luminosity rises and reaches a peak around
$t_{o}\simeq220$. At late times, received $n=0$ photons emitted in
$\mathcal{M}_{1}$ enable the luminosity to stay roughly constant.}%
\label{Fig: phi0-WHSII}%
\end{figure}

In the left panel of FIG. \ref{Fig: phi0-WHSII}, we exhibit the normalized
frequency of photons received by an observer located at $\varphi=0$ and
$\theta=\pi/2$ on the celestial sphere in $\mathcal{M}_{1}$ for the scenario
II. The observed frequency of radially emitted photons with $n=0$ is
represented by the yellow line, which displays three periods. In the first and
last periods, the photons are emitted when the star moves towards and away
from the throat in $\mathcal{M}_{1}$, respectively, and the $n=0$ frequency
line both decreases with the received time; in the intermediate period, the
star emits the photons in $\mathcal{M}_{2}$, and the $n=0$ frequency line
increases. There appears a peak and a dip of the $n=0$ frequency line, which
correspond to the star going through the throat the first time and the second
time, respectively. Similar to the scenario I, the $n=1$ frequency lines
consist of three lines with $b_{1}\gtrsim b_{1}^{\text{ph}}$, $b_{1}\lesssim
b_{1}^{\text{ph}}$ and $b_{2}\simeq b_{2}^{\text{ph}}$, respectively. The
$n=1$ frequency line with $b_{2}\simeq b_{2}^{\text{ph}}$ increases slowly
until the maximum and then decreases rapidly in the first period, rises to a
peak followed by a steep decline in the intermediate period, and gradually
increases in the last period. For the $n=1$ frequency lines with $b_{1}\gtrsim
b_{1}^{\text{ph}}$ and $b_{1}\lesssim b_{1}^{\text{ph}}$, there is a sharp
drop after reaching a peak when the star moves away from the observer, and a
steady increase when the star moves towards the observer. For $n=2$, only two
frequency lines are visible, namely the $b_{1}\lesssim b_{1}^{\text{ph}}$
(orange dots) and $b_{2}\simeq b_{2}^{\text{ph}}$ (brown dots) lines. Note
that the $n=2$ frequency lines are quite similar to the $n=1$ counterparts. In
addition, only the frequency line with $b_{1}\lesssim b_{1}^{\text{ph}}$ is
visible for $n=3$.

The normalized luminosity of the star in the scenario II measured by the
observer is displayed in the right panel of FIG. \ref{Fig: phi0-WHSII}.
Similar to the scenario I, the luminosity decreases slowly before $t_{o}%
\simeq170$, which is dominated by radially emitted photons in the yellow
region. Afterwards, photons emitted in the blue region come to control the
luminosity observation and lead to a flash around $t_{o}\simeq180$.
Subsequently, photons emitted in the yellow region determine the luminosity
observation again and produce a peak around $t_{o}\simeq220$. At late times,
the star travels towards the observer at a large $r_{e}$ in $\mathcal{M}_{1}$,
and hence radially emitted photons would make a dominant contribution to the
total luminosity. In particular, the late-time luminosity remains fairly
constant, which is greatly different from the black hole case.

\section{Conclusions}

\label{sec:CONCLUSIONS}

In this paper, we investigated observational appearances of a point-like
freely-falling star, which emits photons isotropically in its rest frame, in
an asymmetric thin-shell wormhole connecting two spacetimes, $\mathcal{M}_{1}$
and $\mathcal{M}_{2}$. Specifically, two scenarios with different initial
velocities of the star were considered. In the scenario I, the star starts
with a nonzero velocity at spatial infinity of $\mathcal{M}_{1}$ and moves
towards spatial infinity of $\mathcal{M}_{2}$. In the scenario II, the star
falls at rest from spatial infinity of $\mathcal{M}_{1}$, reaches a turning
point in $\mathcal{M}_{2}$ and returns to $\mathcal{M}_{1}$. For the two
scenarios, the frequency distribution and luminosity of the star measured by
all observers and a specific observer on a celestial sphere were obtained by
numerically tracing emitted light rays. Interestingly, it was found that the
absence of the event horizon and the presence of two photon spheres play a
pivotal role in frequency and luminosity observations.

In \cite{Cardoso:2021sip} and \cite{Chen:2022qrw}, observational appearances
of a star freely falling in black holes with one or two photon spheres were
investigated. To compare the wormhole case with the black hole one, we briefly
summarize the main findings of \cite{Cardoso:2021sip,Chen:2022qrw} and this
paper as follows.

\begin{itemize}
\item Black holes with a single photon sphere: The total luminosity of the
star fades out with an exponentially decaying tail, which is determined by
quasinormal modes at the photon sphere. At late times, the specific observer
sees a series of flashes indexed by the orbit number, whose luminosity
decreases exponentially with the orbit number. Moreover, the frequency content
of received photons contains a discrete spectrum of frequency lines indexed by
the orbit number, which decay sharply at late limes.

\item Black holes with double photon spheres: At late times, the total
luminosity first rises to a peak and then decreases with an exponentially
decaying tail. The sub-long-lived quasinormal modes at the outer photon sphere
are responsible for the slowly decaying exponential tail, and the leakage of
photons trapped between the inner and outer photon spheres results in the
luminosity peak. The specific observer sees two series of flashes, which are
mainly determined by photons orbiting outside the outer and inner photon
spheres, respectively. Moreover, the specific observer detects a discrete
spectrum of frequency lines indexed by the orbit number and the photon sphere
that received photons orbit around, which fall steeply at late limes.

\item Wormhole: At late times, the total luminosity first rises to a peak and
then gradually decays with time (scenario I) or remains roughly constant
(scenario II). The luminosity peak is caused by photons travelling between the
two photon spheres (scenario I) or those emitted in $\mathcal{M}_{2}$ nearly
along the radial direction (scenario II). Due to the absence of the event
horizon, a considerable number of photons can still reach observers at late
times, and hence an exponentially decaying tail would not appear. Similarly,
the late-time luminosity measured by the specific observer can be sizable, and
therefore he only sees a bright flash and a faint one (scenario I) or two
bright flashes (scenario II) due to strong background luminance. Moreover, the
specific observer detects frequency lines indexed by the orbit number and the
photon sphere that received photons orbit around. The frequency lines produced
by photons orbiting around the photon sphere in $\mathcal{M}_{1}$ decline sharply (scenario
I) or grow steadily (scenario II) at late limes; those produced by photons
orbiting around the photon sphere in $\mathcal{M}_{2}$ gradually increase at late limes.
\end{itemize}

In short, we showed that the absence of the event horizon in wormholes gives
rise to significantly different optical appearances of a luminous star
accreted onto wormholes at late times. Therefore, these findings can provide
us a novel tool to distinguish wormholes from black holes in future observations.

\begin{acknowledgments}
We are grateful to Guangzhou Guo and Qingyu Gan for useful discussions and
valuable comments. This work is supported in part by NSFC (Grant No. 11875196,
11947225, 12105191, 12275183 and 12275184). Houwen Wu is supported by the
International Visiting Program for Excellent Young Scholars of Sichuan University.
\end{acknowledgments}

\bibliographystyle{unsrturl}
\bibliography{ref}

\end{document}